\documentclass[aps,prb,reprint,superscriptaddress]{revtex4-1}
\newcommand*{\citen}[1]{%
  \begingroup
    \romannumeral-`\x % remove space at the beginning of \setcitestyle
    \setcitestyle{numbers}%
    \cite{#1}%
  \endgroup   
}
\usepackage{graphicx}
\newcommand{\cmmnt}[1]{}
\begin{document}

%\preprint{}

\title{Glass-Like Thermal Conductivity in Nanostructures of a Complex Anisotropic Crystal}


\author{Annie Weathers}
\affiliation{Department of Mechanical Engineering, The University of Texas at Austin, Austin, Texas 78712, United States}
\author{Jes{\'u}s Carrete}
\affiliation{Institute of Materials Chemistry, Technische Universität Wien, A-1060 Vienna, Austria}
\author{John P. Degrave}
\affiliation{Department of Chemistry, The University of Wisconsin - Madison, Madison, Wisconsin, 53706, United States}
\author{Jeremy M. Higgins}
\affiliation{Department of Chemistry, The University of Wisconsin - Madison, Madison, Wisconsin, 53706, United States}
\author{Arden L. Moore}
\affiliation{Mechanical Engineering Department, Louisiana Tech University, Ruston, LA 71272, USA}
\author{Jaehyun Kim}
\affiliation{Department of Mechanical Engineering, The University of Texas at Austin, Austin, Texas 78712, United States}
\author{Natalio Mingo}
\affiliation{Laboratoire d’Innovation pour les Technologies des Energies Nouvelles et les Nanomatériaux, Commissariat à l’Énergie Atomique Grenoble, Grenoble 38054, France}
\author{Song Jin}
\affiliation{Department of Chemistry, The University of Wisconsin - Madison, Madison, Wisconsin, 53706, United States}
\author{Li Shi}
\affiliation{Department of Mechanical Engineering, The University of Texas at Austin, Austin, Texas 78712, United States}

\email[]{lishi@mail.utexas.edu}
%\thanks{}


\date{\today}

\begin{abstract}
Size effects on vibrational modes in complex crystals remain largely unexplored, despite their importance in a variety of electronic and energy conversion technologies. Enabled by advances in a four-probe thermal transport measurement method, we report the observation of glass-like thermal conductivity in $\sim$20 nm thick single crystalline ribbons of higher manganese silicide, a complex, anisotropic crystal with a $\sim$10 nm scale lattice constant along the incommensurate \emph{c} axis. The boundary scattering effect is strong for many vibrational modes because of a strong anisotropy in their group velocities or diffusive nature, while confinement effects are pronounced for acoustic modes with long wavelengths along the \emph{c} axis. Furthermore, the transport of the non-propagating, diffusive modes is suppressed in the nanostructures by the increased incommensurability between the two substructures as a result of the unusual composition of the nanostructure samples. These unique effects point to diverse, new approaches to suppressing the lattice thermal conductivity in complex materials.
\end{abstract}

%\pacs{}

\maketitle

\section{Introduction}

Atomic vibrations in a solid are intimately coupled to the excitations of the electronic and spin degrees of freedom, and influence not only thermal but also electronic, optical, and magnetic properties of materials. In simple periodic crystals, the vibrational modes can be successfully treated as extended, propagating phonon modes, and the thermal conductivity contribution from these propagating modes can be calculated from first principles and numerical solutions to the Peierls-Boltzmann transport equation without the use of fitting parameters \cite{Lindsay2016,Broido2013}. In comparison, thermal transport in amorphous solids has been explained by a number of theories of heat transport by non-propagating modes, which contribute to the heat current either through diffusive random walks \cite{Cahill1992,Allen1993,He2011} or through anharmonic coupling with propagating modes \cite{Nakayama1999}.

Despite these theoretical advances, there remain a number of important questions on thermal transport by vibrational modes in solids. In particular, recent progress in experimental methods has allowed direct thermal transport measurements of individual nanostructures with a characteristic size comparable to the mean free path or even the wavelength of the vibrational modes in a crystal \cite{Weathers2013a}. Such measurements have revealed size-dependent thermal transport properties in a number of crystalline nanostructures, including experimental results that cannot be explained by prior theories \cite{Wingert2016,Li2003,Hochbaum2008,Wingert2015}. Among the notable examples, the measured thermal conductivities of Si nanowires become considerably lower than the calculated Casimir limit based on diffuse surface scattering of phonons when the diameter is reduced below about 20 nm or when the surface is rough \cite{Li2003,Hochbaum2008}. The unusually low thermal conductivity found in these crystalline Si nanostructures is desirable for thermal insulation, and could be beneficial for thermoelectric materials if the electronic mobility is not suppressed considerably in the nanostructures. However, the exact cause of such low thermal conductivity has remained unclear. Reduced phonon group velocities and wave interference effects such as coherent surface roughness scattering have been considered \cite{Sadhu2012,Martin2009,Zhu2016,Chen2011}. Meanwhile, other semi-classical effects such as phonon scattering by high-concentration interior defects \cite{Cahill2014} and backscattering by rough surfaces \cite{Moore2008} have been investigated. 

Besides these perplexing size effects on the propagating modes in crystalline Si nanostructures \cite{Shi2012}, the thermal conductivity of amorphous Si nanostructures was found to decrease considerably with decreasing thickness \cite{He2011,Kwon2017}. This size dependence has suggested the important role of propagating phonon modes with long mean free paths even in amorphous Si, whereas the weakly-localized diffusive modes were assumed to be unaffected by the size reduction.

In addition to simple periodic crystals and entirely disordered systems such as crystalline and amorphous silicon, there exists a variety of complex crystals that are characterized by the coexistence of order and disorder at different length scales. Many complex crystals exhibit unusual thermal, electric, optoelectronic, and magnetic properties that are influenced by the lattice dynamics. Due to the presence of a large number of atoms in the unit cell of a complex crystal, numerous non-propagating modes coexist with propagating modes. In addition, the lattice constants of a complex crystal can be one or two orders of magnitude larger than the atomic scale lattice constant in a simple crystal such as Si, and can become comparable to the critical dimensions of nanostructures that can be synthesized. Thus, the effect of confinement on a propagating mode in a complex crystal can potentially become very pronounced compared to the situation in silicon. However, the size effects on both the propagating and non-propagating modes in complex crystals have remained largely unexplored. 

In this article, we report a combined experimental and theoretical study of the size effects on thermal transport in $\sim$20 nm thick ribbon structures of a representative complex crystal, higher manganese silicide (HMS), which is one of the leading thermoelectric materials made from earth-abundant, non-toxic elements \cite{Akio2016,Chen2015a}. Because of an advance in making clean electrical contact to suspended nanostructures, we are able to conduct four-probe thermoelectric measurements of the suspended nanoribbon and obtain the intrinsic thermal conductivity, which is considerably suppressed compared to the bulk values. Remarkably, both the magnitude and temperature dependence of the obtained intrinsic thermal conductivity of the single-crystalline HMS nanoribbon samples resemble those of \emph{amorphous} silica glass. Theoretical calculations attribute the finding to several unique size effects that influence the vibrational modes in the complex, anisotropic crystal, and suggest that the diffusivity of non-propagating modes can be suppressed by increasing the incommensurability between the sublattices in the complex crystals. 

\section{Experimental Methods and Results}
Belonging to the family of Nowotny chimney ladder phases, the complex HMS structure consists of a $\beta$-Sn tetragonal sublattice of Mn atoms surrounding coupled helices of Si aligned along the \emph{c} axis  \cite{deRidder1971,Ye1986}, as illustrated in Fig. 1. The hard Mn sublattice maintains a nearly constant \emph{c} axis unit cell length of $c_{Mn}$ = 4.3 $\AA$ across all phases, while the relatively soft Si sublattice length, $c_{Si}$, varies slightly depending on the stoichiometric composition \cite{deRidder1971,Ye1986,Higgins2008}. Electron microscopy and diffraction studies have further suggested that these idealized commensurate phases are often a result of averaging over a large sampling area, and that only incommensurate structures exist in real crystals \cite{deRidder1976}. Indeed, HMS is often described within the context of aperiodic crystals due to the incommensurate nature of the sublattices and the resulting structural complexity \cite{deBoissieu2008}. In addition, bulk HMS crystals grown from a melt are often synthesized with unintentional inclusions of multiple HMS phases in addition to the metallic B20 MnSi phase, which precipitates perpendicular to the \emph{c} axis as a result of the peritectic decomposition of the HMS phase \cite{Levinson1973,Girard2014}. It has remained an outstanding question whether the varying incommensurability of different HMS phases can lead to differences in the thermal conductivity. 
\begin{figure}
\includegraphics{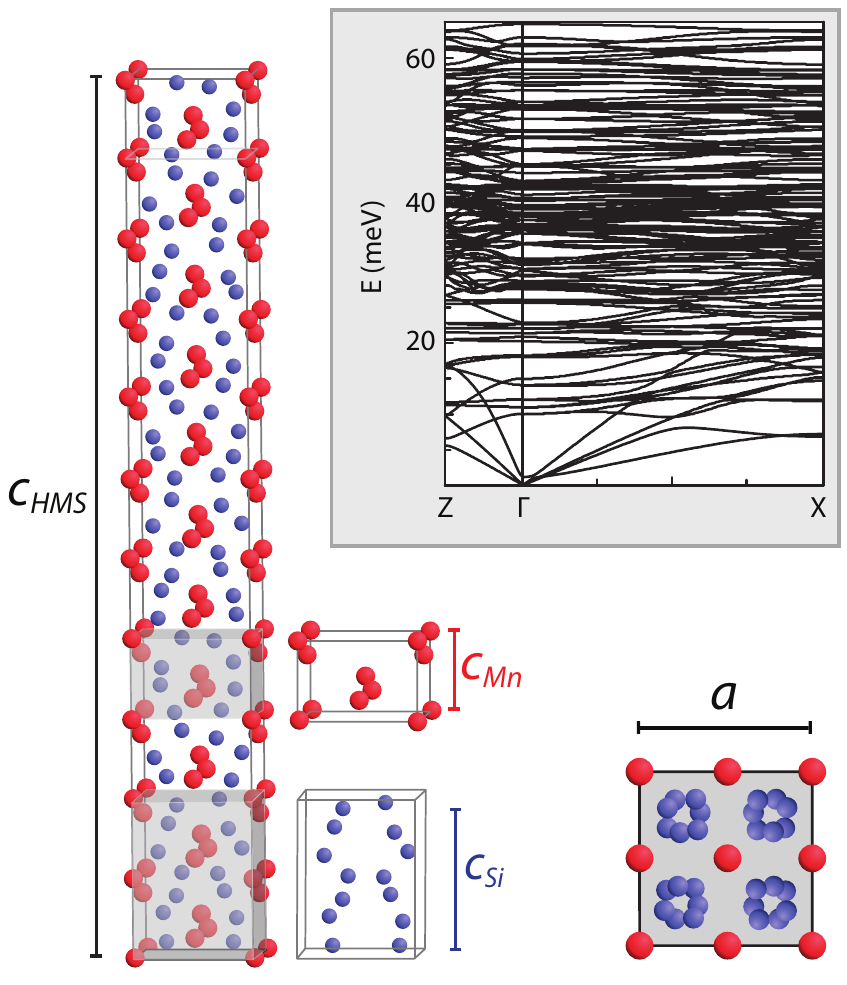}
\caption{A representative crystal structure and vibrational spectrum of HMS. The Mn atoms (red) form a tetragonal lattice, while the Si atoms (blue) form coupled helices aligned along the \emph{c} axis to form a homologous series of compounds. The c lattice parameter of the specific Mn$_{11}$Si$_{19}$ phase shown here, $c_{HMS}$, is 47.7 $\AA$, while the a lattice parameter, $a_{HMS}$, is 5.5 $\AA$. The top right inset shows the calculated vibrational spectrum of the Mn$_4$Si$_7$ phase from ref. 29.}
\end{figure}

In this work, HMS nanoribbons \cite{Higgins2008} without the MnSi phase are used for studying thermal transport in complex crystals. The HMS nanoribbons were grown by chemical vapor deposition (CVD) at a growth temperature of $700$ $^{\circ}$C \cite{Higgins2008}. Because the HMS nanostructures are coated with native oxide, it is not possible to make electrical contact by directly placing the sample on the electrodes of the suspended device used for the thermal transport measurements. This challenge has been overcome by developing a method of transferring the sample together with four Pd contact pads onto the suspended device, so that four-probe measurements of the intrinsic thermal and thermoelectric transport properties can be made to the nanoribbon (NR) samples, as described in the Supplemental Material. Figure 2 shows scanning electron microscopy (SEM) images of a HMS nanoribbon sample (NR1) suspended across the measurement device with false coloring (blue) of the Pd contact pads, which have been transferred to the device together with the NR sample. 

Both samples, referred to as NR1 and NR2 hereafter, were found to be NRs with widths of $195 \pm 3$ and $95 \pm 5$ nm and thicknesses of $24 \pm 4$ and $28 \pm 3$ nm, respectively (Supplemental Material). Figures 3a-b show the transmission electron microscopy (TEM) images of NR1, with the [001] crystallographic direction indicated, which is oriented parallel to the silicon helices. The results clearly reveal the single crystalline nature of the nanostructure sample. The contrast modulation bands observed in Fig. 3a are associated with the mismatch between the Mn and Si sublattices along the \emph{c} axis \cite{Higgins2008,Chen2015}. The electron diffraction pattern from NR1 (Fig. 3b) displays bright central peaks associated with the Mn tetragonal sublattice, and closely spaced satellite peaks along the \emph{c} axis associated with the Si sublattice. The TEM images and diffraction patterns are used to determine the crystallographic direction of NR1 and NR2 along the NR transport direction, which are found to be at angle ($\theta$) of $32^{\circ}$ and $26^{\circ}$ from the \emph{c} axis, respectively. The thickness of the native oxide is found to be less than 4 nm on the lateral sides of the NRs. 
\begin{figure}
\includegraphics{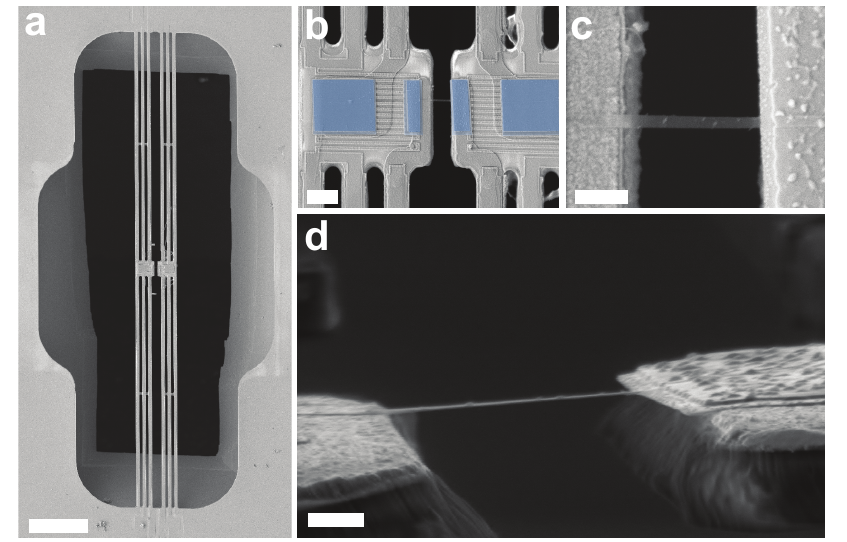}
\caption{Electron microscopy images of the nanostructure on the measurement device. (a) Scanning electron microscopy (SEM) images of a suspended measurement device. (b-c) SEM images of HMS NR1 with false coloring (blue) of the Pd contact pads transferred to the device with the sample. (d) 85$^\circ$ tilted SEM image of NR1. Scale bars are (a) 50 $\mu$m, (b) 3 $\mu$m, (c) 1 $\mu$m, and (d) 500 nm.}
\end{figure}

\begin{figure}
\includegraphics{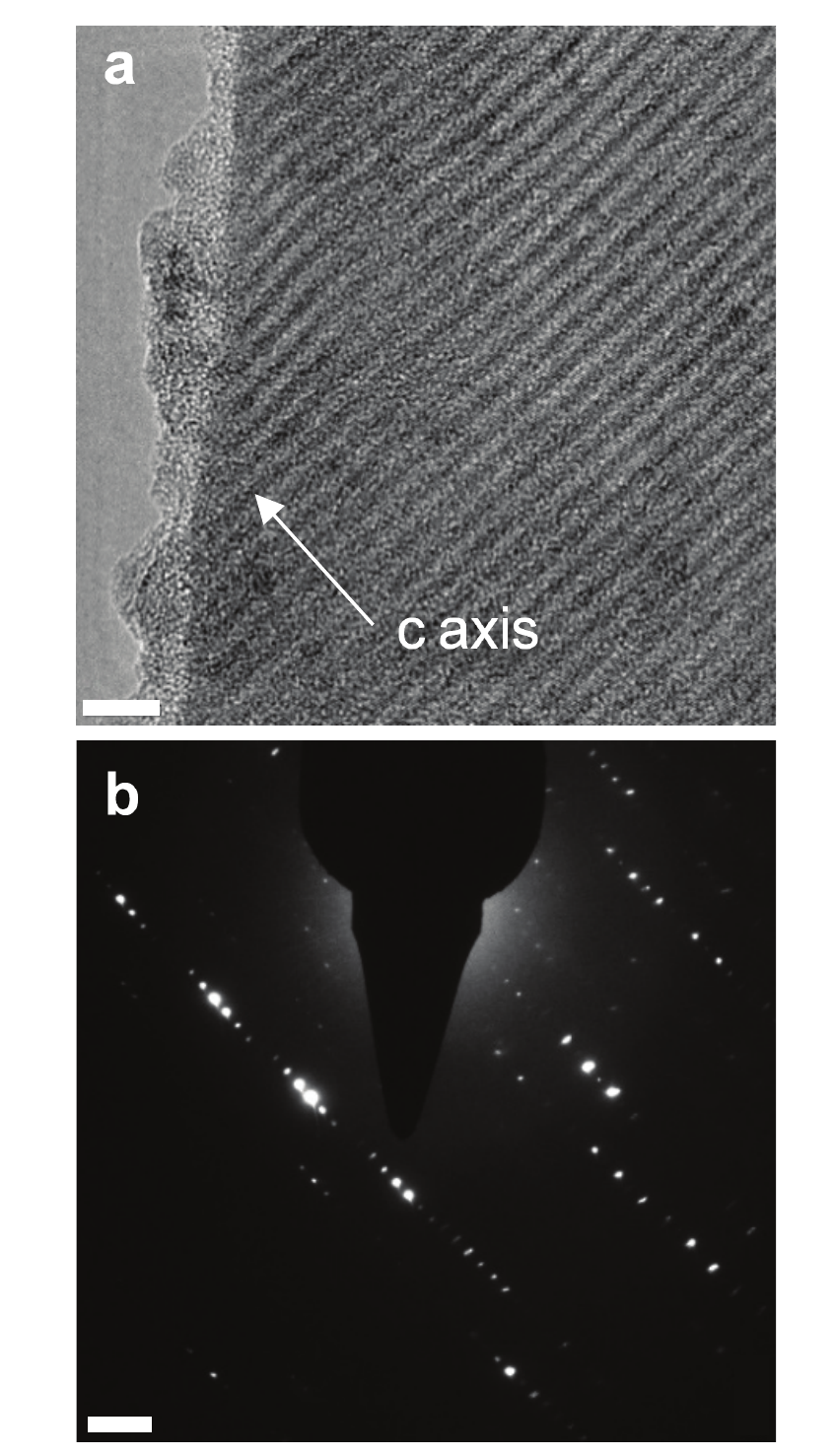}
\caption{Transmission electron microscopy (TEM) analysis of the HMS nanoribbon samples. (a) High resolution TEM (HRTEM) images of NR1. (b) Electron diffraction pattern of NR1 showing the closely spaced satellite peaks associated with the \emph{c} axis configuration of the Si sublattice. Scale bars are 5 nm in (a) and 2 nm$^{-1}$ in (b).}
\end{figure}

The thermal conductance of the HMS sample was as low as $1.85 \times 10^{-9}$ W K$^{-1}$ and $0.73 \times 10^{-9}$ W K$^{-1}$ at room temperature for NR1 and NR2, respectively. Therefore, the low thermal conductance samples were measured with a sensitive differential background subtraction method  \cite{Weathers2013}. In addition, the thermal contact resistance determined from a four-probe thermoelectric measurement \cite{Mavrokefalos2007} was found to increase with decreasing temperature and ranged from 6 - 23\% of the measured total thermal resistance of NR1, and negligible for NR2 (Supplemental Material).

The measured thermal conductivities of NR1 and NR2 are shown in Fig. 4a, together with the literature data for bulk HMS crystals \cite{Chen2015}. The four-probe thermal conductivity results for NR1 and NR2 are comparable to the highest values that were measured for a number of other HMS NR and nanowire samples with the use of a two-probe thermal measurement approach, due to the lack of electrical contact to the sample, or a four-probe method where the electrical contact was made using focused electron beam assisted metal deposition \cite{Moore2010}. In Fig. 4a, the circular symbols represent the effective thermal conductivity of the nanoribbons, which consists of both the HMS core and the amorphous native-oxide shell. The upper limit of the thermal conductivity of just the HMS core ($\kappa_{NR}$) is calculated from the measured effective thermal conductivity by assuming the thickness of the amorphous oxide is 4 nm on all side walls with a thermal conductivity taken to be that of thin silicon oxide grown by plasma enhanced chemical vapor deposition \cite{Lee1997}. Indeed, the native oxide on the HMS nanowires and NRs has been shown previously to be composed primarily of SiO$_x$ \cite{Higgins2008}. The as-obtained maximum values of $\kappa_{NR}$ are shown as the upper limit to the shaded regions in Fig. 4a. The reported bulk thermal conductivities along the \emph{a} axis ($\kappa_a$) and \emph{c} axis ($\kappa_c$) are used to calculate the bulk thermal conductivity ($\kappa_\theta$) along the crystal direction corresponding to the transport directions of the two nanoribbons, as shown as the upper and lower limits of the grey shaded area of Fig. 4a. The thermal conductivities of the NRs increase with temperature, and are approximately a factor of 2.5 times lower than the bulk $\kappa_\theta$ value at room temperature. Moreover, it is remarkable that the suppressed lattice thermal conductivity of the \emph{single-crystalline} HMS nanostructures is comparable to that of \emph{amorphous} fused silica glass in both temperature dependence and magnitude (green curve, Fig. 4a).

\begin{figure*}
\includegraphics{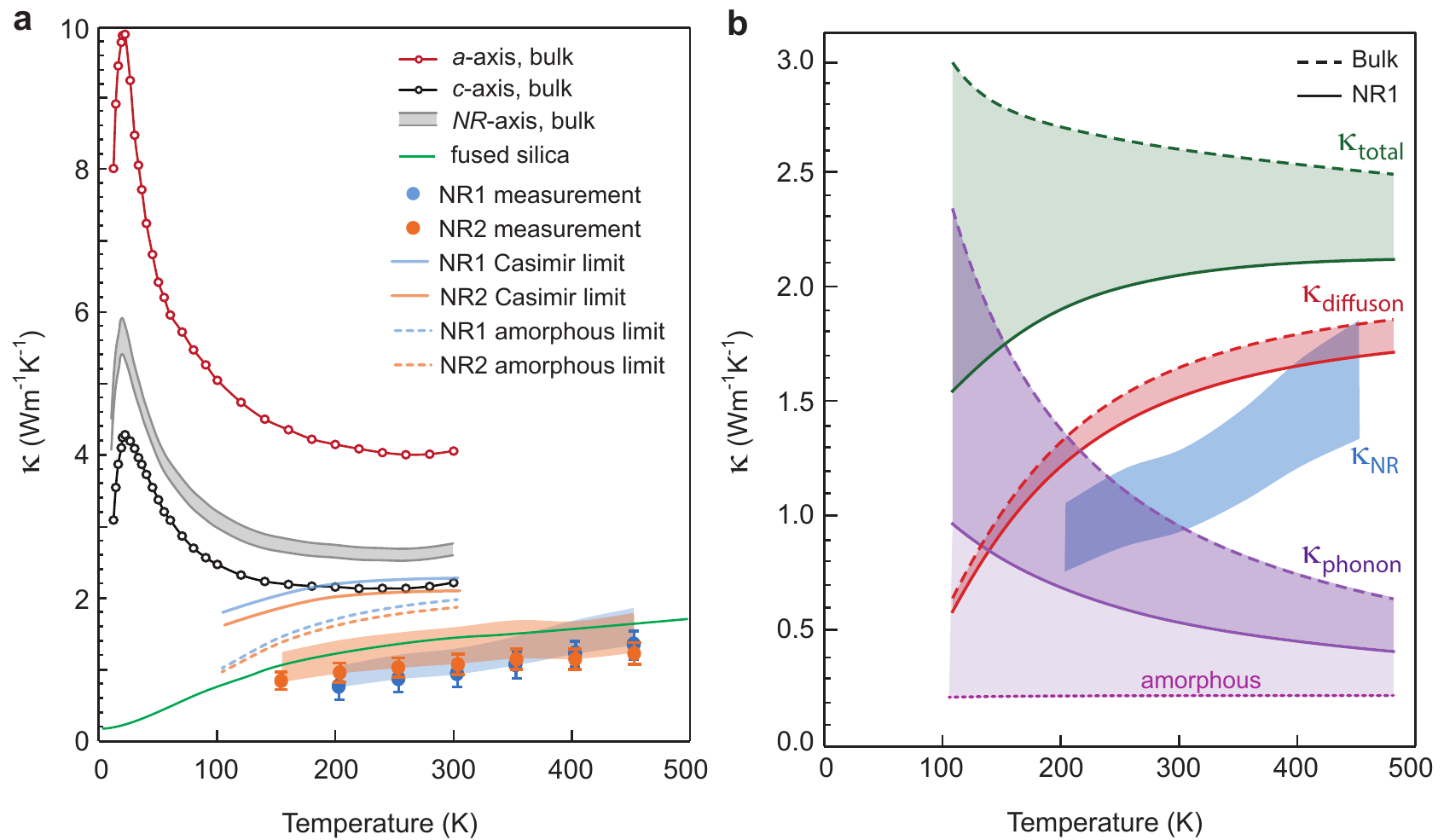}
\caption{(a) Measured effective thermal conductivity of NR1 (blue filled circles) and NR2 (orange filled circles). The error bars are dominated by the error in the NR thickness measurement. The maximum thermal conductivity of the HMS core, $\kappa_{NR}$, assuming a SiO$_2$ shell thermal conductivity from ref. 33\cmmnt{\cite{Lee1997}}, is shown as the upper limit to the shaded region. The thermal conductivity of bulk crystalline HMS from ref. 29\cmmnt{\cite{Chen2015}} is shown as red and black unfilled circles along the \emph{a} axis and \emph{c} axis, respectively. The bulk values along $\theta  = 32^{\circ}$ and $26^{\circ}$, which are the transport directions of the two NRs, are shown as the upper and lower limits of the gray shaded region. The thermal conductivity of fused silica glass is shown as the green line for comparison \cite{Touloukian1971}. The solid and dashed blue and orange lines are the calculated Casimir limit and amorphous limit for the two NR samples according to the approaches described in the text. (b)  Contributions to the thermal conductivity from phonons (purple) and diffusons (red) for both NR1 (solid lines) and for the bulk crystal (dashed lines). The reduction in the thermal conductivities in the nanostructured samples are shown as the shaded areas between the dashed and solid lines. The experimental thermal conductivity data for NR1 is shown as the shaded blue area. The calculated amorphous limit to the phonon contribution is indicated by the bottom dotted line.}
\end{figure*}

\section{Theoretical Analysis}
To explain the unusually low thermal conductivity of the HMS NRs we consider a number of possible reasons as discussed below. We find that diffuse surface scattering of propagating modes and confinement of both the propagating and non-propagating modes are insufficient in explaining the low thermal conductivity, but that suppressed diffusivity of the non-propagating modes as a result of increasing incommensurability in the HMS NRs is an important cause of the large reduction in thermal conductivity.

\subsection{Lattice Dynamics Model for Bulk HMS}
Because the electronic contribution to the thermal conductivity of the NR samples is negligible compared to the lattice contribution in the temperature range of the measurements (Supplemental Material), the observed low thermal conductivity is due entirely to a modification in the lattice thermal conductivity. A recent lattice dynamics model was able to explain the thermal conductivity of bulk HMS based on the combined contributions from both propagating phonon modes (propagons) and diffusive modes (diffusons) with energies below and above 20 meV, respectively \cite{Chen2015}. This cutoff energy was chosen based on inelastic neutron scattering (INS) measurements and lattice dynamics calculations, which observed vibrational modes with clearly defined group velocities at energies below 20 meV (Fig. 1 inset) \cite{Chen2015}. At energies higher than 20 meV, most of the observed modes exhibited very broad linewidths that are characteristic of diffusive modes, with the exception of some modes with wave vectors perpendicular to the \emph{c} axis that appeared to possess non-vanishing velocities. At energies below 7 meV, the average group velocity along the \emph{c} axis was found to be considerably higher than that along the \emph{a} axis of HMS because of stronger atomic bonding along each Si helical ladder compared to the bonding between adjacent ladders and the Mn chimney sublattice.

\subsection{Diffuse Surface Scattering}
The measured thermal conductivity decreases with decreasing temperature from 450 K to 150 K. Because the propagating modes increase the thermal conductivity of bulk HMS with decreasing temperature from about 250 K to 50 K, the observed temperature dependence for the HMS NR samples is indicative of a suppressed contribution from propagating modes by diffuse surface scattering. Using the group velocities of the vibration modes determined in the previous model \cite{Chen2015}, we first calculated the boundary scattering mean free paths ($\Lambda_{b,\alpha}$) of the propagating modes with energies below 20 meV. In an isotropic crystal, the boundary scattering mean free path of a thin film is approximately the thickness of the film. However, in an anisotropic nanostructured sample an effect known as phonon focusing causes an effective focusing of energy along the direction of highest group velocity \cite{McCurdy1970}. In the HMS NRs, this focusing effect results in $\Lambda_{b,\alpha}$ being larger than the value in an isotropic crystal when the fast axis is close to the NR axis, i.e., $E<10$ meV. The focusing effect is reversed for $E > 10$ mV, when the disordered Si ladder structure results in a lower group velocity along the incommensurate \emph{c} axis, so that $\Lambda_{b,\alpha}$ becomes smaller than the corresponding value of an isotropic crystal. With these two opposite focusing effects accounted for in our model, the calculated Casimir limit of the contribution from propagons with $E < 20$ meV in the NRs is about 40\% of the bulk value at 300 K, as shown by the purple curves in Fig. 4b.   

The thermal conductivity increases very gradually with temperature in the experiment temperature range. This trend is similar to that of amorphous glass, and indicates a dominant contribution from non-propagating modes. The thermal conductivity contribution from modes above the 20 meV cutoff was calculated according to the diffuson thermal conductivity expression (Supplemental Material) \cite{Allen1993}. In principle, the presence of sample boundaries should have negligible effect on weakly localized diffuson modes. However, a peculiar situation for HMS is the presence of modes above 20 meV with non-zero group velocity components along the commensurate \emph{a} axis, while almost all the modes along the incommensurate \emph{c} axis have nearly zero group velocity. As a result, there is a similar focusing of the high energy modes toward the boundaries of the sample. If this focusing effect is also applied to the model of these higher energy modes, this approximation can lead to a noticeable reduction in the effective diffusivity ($D$). This boundary treatment leads to a 15\% reduction in the diffuson thermal conductivity at 300 K compared to the bulk value, as shown in Fig. 4b. However, even when both propagating modes and higher energy modes are treated with essentially the Casimir model of diffuse boundary scattering, the calculated Casimir limit is still considerably higher than the measurement results of the two HMS NR samples, as shown by the solid orange and blue lines in Fig. 4a. Thus, we must to consider other possible causes for the unusually low thermal conductivity of HMS NRs. 

\subsection{Confinement of Vibrational Modes}
A number of prior reports have suggested that the phonon group velocity can be reduced considerably compared to the bulk value when the diameter of a Si nanowire is in the sub-20 nm regime \cite{Wingert2015,Zhu2016,Zou2001}, so that the propagon contribution falls below the Casimir limit. In some works, the reduction is attributed to a decreased elastic modulus \cite{Wingert2015}. In comparison, atomistic and first-principles calculations have found that the elastic properties of silicon nanostructures can only be reduced appreciably by the surface effect alone when the characteristic size is reduced well below 10 nm \cite{Mingo2003}. It has been suggested that surface oxide and internal defects, instead of surface stress, could have resulted in the considerable suppression of the elastic constant in nanowires thicker than 10 nm \cite{Sadeghian2010}. Grain boundaries and defects can also be expected and may vary with size in polycrystalline Si nanotubes measured in a recent work \cite{Wingert2015}. In addition, it has been pointed out that the group velocity along the nanowire axis should be compared with the group velocity component ($v_x$) along the same crystalline direction in the bulk crystal, instead of the total group velocity magnitude ($v$) \cite{Shi2012,Mingo2003,Li2012}. Hence, the size effects on the elastic modulus and phonon group velocity in $\sim$20 nm Si nanostructures remain to be better understood. 

Compared to the Si nanostructures, the HMS NRs are unique due to the large lattice constant along the \emph{c} axis, which can exceed 10 nm. In theory, the minimum wavelength of pure acoustic modes along the \emph{c} axis is twice the \emph{c} lattice constant. For a mode with non-zero group velocity components both perpendicular ($v_\perp$) and parallel ($v_{//}$) to the top and bottom surfaces, the spatial extent of the mode would be limited by the HMS core thickness ($t_c$) to be on the order of $t_c v_{//}/v_\perp$, which is also on the $\sim$20 nm scale and comparable to the minimum allowable wavelength along the \emph{c} axis. Hence, the confinement effect can reduce the group velocity of these long-wavelength acoustic modes.  At the same time, there exist very low-lying optical modes with large group velocities in incommensurate crystals where the interaction between the two sublattices is weak \cite{Chen2016,Hastings1977,Axe1982}. The low energy of these pseudo-acoustic modes originates from the small energy cost of sliding or twisting one sub-lattice against another, nearly stationary, sublattice. As shown in prior INS measurements of HMS \cite{deBoissieu2008,Currat2002}, these sliding or twisting modes approximately follow the short periodicity of the sub-lattice, instead of the long lattice constant of the whole structure. Hence, the size confinement effect is expected to be weaker for these sub-lattice pseudo-acoustic modes than for the pure acoustic modes of the entire unit cell. 

Nevertheless, to investigate whether confinement or localization of the long-wavelength modes can reduce the thermal conductivity to the measurement results, the thermal conductivity contribution from modes with energy below 20 meV was calculated as that of diffusive modes with a diffusivity given by $v_{\alpha,\theta}^2\pi/\omega$, consistent with the so-called minimum thermal conductivity model of an amorphous material \cite{Cahill1992}. When the thermal conductivity contribution from modes above 20 meV is still assumed to be the same as the red solid line of Fig. 4b, the as-calculated amorphous limit of the NR thermal conductivity, shown as the dashed lines in Fig. 4a, is still higher than the measurement results. Even more importantly, the calculated diffuson contribution alone is still higher than the measured thermal conductivity even when the focusing effect is accounted for, as shown in Fig. 4b.

 This discrepancy suggests that the contribution from modes higher than 20 meV in the HMS NRs must be lower than the diffuson contribution calculated for the bulk crystal. Indeed, the measurement results can be matched with the calculation results when the contribution from diffusive modes above 20 meV is reduced by a factor of 3 compared to solid line of Fig. 4b and the contribution from the propagating modes below 20 meV is taken to be the Casimir limit. 

\subsection{Spatial Confinement of Weakly Localized Modes}
An important question in the study of vibrational modes in complex crystals is whether the spatial confinement of a diffusive mode can considerably reduce its diffusivity. For example, the eigenfunctions of diffusive modes in amorphous silicon have been shown to have a polynomial spatial decay \cite{Allen1999,He2011a}, suggesting a weak localization and possible sensitivity to adjacent boundaries. Analogous to the important effect of volume confinement on the percolation threshold of nanocomposites and disordered media \cite{Stevens2008}, a network of weakly-localized diffuson modes would likely be affected by the sample boundaries. Additionally, as discussed in a recent molecular dynamics study of two-dimensional amorphous graphene and one-dimensional diamond nanothreads \cite{Zhu2016a}, the thermal conductivity contribution from diffusive modes can be suppressed in low-dimensional systems, because of an increased chance of returning to their starting point in a recurrent random walk. However, the HMS nanoribbon samples are still three-dimensional (3D) structures. In addition, the radius of diffusion can be calculated as $r_d=\pi/\sqrt{D / 2 \omega}$ (ref.  \citen{Beltukov2013}), which is less than 1 nm based on the $D$ values obtained for HMS. Because the thickness is still much larger than $r_d$, the surface is not expected to suppress the diffusivity considerably.     

\subsection{Effect of Incommensurability on Thermal Conductivity}
While surface scattering and confinement cannot explain the low thermal conductivity of the samples, an important question is whether the diffusivity of non-propagating modes can vary considerably in different HMS phases because of varying incommensurability between the sublattices. The measured thermal properties of bulk HMS crystals are often an average over multiple phases present in the crystal. In comparison, TEM measurements along the NRs have revealed that the NRs are clearly single crystalline and lack the MnSi inclusions found in bulk HMS crystals. The MnSi inclusions have been described as soliton walls that effectively separate deformed commensurate phases as a result of the relaxation of the soft Si sublattice \cite{Zaitsev1995}. The lack of a soliton structure in these NRs is a clear indication that such relaxation of the disorder in the Si sublattice has not occurred due to the relatively low growth temperature ($700$ $^{\circ}$C) of the NRs, compared to $900$ $^{\circ}$C for the growth of HMS bulk crystals. Without this relaxation, the average crystal structure of the short NRs should contain a higher degree of incommensurability than the much larger bulk HMS crystals. 

The experimental results obtained for the NRs suggest that the increasing degree of incommensurability in the complex structures could result in a further reduction of the diffusivity via an increase in the unit cell length. While it is known that increasing disorder results in decreasing thermal conductivity, the effect of incommensurability on the thermal conductivity has remained elusive for not only HMS but also other complex crystals. A complete understanding of the effect of incommensurability would require the development of new theoretical capabilities for  modeling thermal transport in the complex crystal without adjustable parameters,  which represents a new direction for providing theoretical guidance to manipulate the thermal properties of complex structures. The experimental and analytical results presented here suggest that such manipulation can lead to apparent results. 

\section{Summary}
The advances in the four-probe thermal transport measurement method has allowed us to establish that the thermal conductivity of $\sim$20 nm thick crystalline ribbons of the complex crystal HMS is as low as that of amorphous silica glass. Besides a reversed focusing effect that leads to a short boundary scattering mean free path for many modes with the higher group velocity component pointing to the boundary, the wavelength of pure acoustic modes for the whole structure is long along the \emph{c} axis so that these modes are strongly confined in the $\sim$20 nm nanostructure. However, these two unique effects associated with the complex anisotropic crystal structure are insufficient to explain the remarkable glass-like thermal conductivity, which is lower than not only the calculated Casimir limit based on diffuse surface scattering, but also the calculated amorphous limit where only diffusive modes are present. The diffusivities of the diffusive modes must have been suppressed considerably in the HMS NRs compared to their important contribution in the bulk. Although it is generally known that increasing disorders should lead to a decreasing thermal conductivity, it has been a question whether different HMS phases can have different thermal conductivity, or more generally, whether the thermal conductivity and mode diffusivity of complex, aperiodic crystals can be tuned by varying the degree of incommensurability. This question cannot be answered by measuring bulk HMS crystals with multiple phases, but has been addressed here by the single NR measurements with the enhanced four-probe thermoelectric transport measurement capability, which is expected to be widely applicable for studying size effects on thermal transport in both complex and simple crystals. In conjunction with the pronounced size effects on many vibrational modes in the complex crystal with strong anisotropies in the group velocities or diffusive nature, increasing incommensurability can be an effective approach to suppressing the diffuson contribution and thermal conductivity.   

\begin{acknowledgments}
The authors thank Drs. Feng Zhou, Ankit Pokhrel, and Xi Chen for helpful discussions, and Mr. Matthew Stolt for his assistance with sample preparation. The work is primarily supported by the US National Science Foundation (NSF)  Department of Energy (DOE) Joint Thermoelectric Partnership (NSF award numbers: CBET-1048767 and CBET-1048625) and NSF Thermal Transport Program (CBET-1336968 and CBET-0933454). A.W. is supported by a NSF Graduate Research Fellowship. J.C. and N.M. acknowledge support from project Carnot SIEVE.
\end{acknowledgments}

\end{document}